\documentclass[11pt]{article}
\usepackage[margin=1in]{geometry}
\usepackage{amsmath}
\usepackage{amssymb}
\usepackage{amsthm}
\usepackage{amsfonts}
\usepackage{physics}
\usepackage{graphicx}
\usepackage{dcolumn}
\usepackage{bm}
\usepackage{authblk}

\usepackage[square,numbers,sort&compress]{natbib}
\title{Analyses of protein cores reveal fundamental differences between solution and crystal structures}

\author[1,2]{Zhe Mei}
\author[1,3]{John D. Treado}
\author[1,4]{Alex T. Grigas}
\author[5,6]{Zachary A. Levine}
\author[7]{Lynne Regan}
\author[1,3,4,8,9]{Corey S. O'Hern}

\affil[1]{Integrated Graduate Program in Physical \& Engineering Biology, Yale University, New Haven, Connecticut 06520, USA}
\affil[2]{Department of Chemistry, Yale University, New Haven, Connecticut 06520, USA}
\affil[3]{Department of Mechanical Engineering \& Materials Science, Yale University, New Haven, Connecticut 06520, USA}
\affil[4]{Graduate Program in Computational Biology \& Bioinformatics, Yale University, New Haven, Connecticut 06520 USA}
\affil[5]{Department of Pathology, Yale University, New Haven, Connecticut 06520, USA}
\affil[6]{Department of Molecular Biophysics and Biochemistry, Yale University, New Haven, Connecticut, 06520}
\affil[7]{Institute of Quantitative Biology, Biochemistry and Biotechnology, Center for Synthetic and Systems Biology, School of Biological Sciences, University of Edinburgh}
\affil[8]{Department of Physics, Yale University, New Haven, Connecticut 06520, USA}
\affil[9]{Department of Applied Physics, Yale University, New Haven, Connecticut 06520, USA}

\begin{document}

\maketitle

\begin{abstract}
There have been several studies suggesting that protein structures
solved by NMR spectroscopy and x-ray crystallography show significant
differences. To understand the origin of these differences, we
assembled a database of high-quality protein structures solved by both
methods. We also find significant differences between NMR and crystal
structures---in the root-mean-square deviations of the C$_{\alpha}$
atomic positions, identities of core amino acids, backbone and
sidechain dihedral angles, and packing fraction of core
residues. In contrast to prior studies, we identify the physical basis
for these differences by modelling protein cores as jammed packings of
amino-acid-shaped particles. We find that we can tune the jammed
packing fraction by varying the degree of thermalization used to
generate the packings. For an athermal protocol, we find that the
average jammed packing fraction is identical to that observed in the cores of
protein structures solved by x-ray crystallography. In contrast,
highly thermalized packing-generation protocols yield jammed packing
fractions that are even higher than those observed in NMR structures.
These results indicate that thermalized systems can pack more densely
than athermal systems, which suggests a physical basis for the
structural differences between protein structures solved by NMR and
x-ray crystallography.
\end{abstract}

It is generally acknowledged that protein structures determined by
x-ray crystallography versus NMR exhibit small but significant
differences. It is by no means resolved, however, whether these
differences stem from differences in the experimental methods
themselves, or if they reflect physical differences in proteins under
the different conditions in which the measurements are
made~\citep{comp:GarbuzynskiyPSFB2005,comp:SchneiderPSFB2009,comp:MaoJACS2014,comp:LemanPSFB2018,comp:BestPNAS2006,comp:YangStructure2007,comp:CarugoOpenBiochemistry2010,comp:EverettProteinScience2016}. To
begin to answer this question, one must directly compare high-quality
structures of the same protein solved by both methods. Choosing x-ray crystal structures based on their resolution is a straightforward way to identify well-specified structures. In our database of structures solved by both x-ray crystallography and NMR, we only include structures that have been solved by x-ray crystallography at a resolution of 2\AA~ or less. 

There is, however, no universally accepted metric to assess the
quality of NMR structures. We therefore defined one. Specifically, we
determined the number of NMR restraints per residue beyond which
structures do not change significantly with the addition of more
restraints, and only used structures with at least this number of
restraints per residue on average. Applying these selection criteria,
we created a data set of $16$ proteins whose structures have been
determined by both x-ray crystallography and NMR. We created an
additional dataset of $51$ high-quality NMR protein structures
(defined in the same way), for which there is no companion x-ray
crystal structure, in an attempt to exclude any influence of
‘crystallizability’ on the NMR protein structures. In addition, as a
reference set of high-resolution protein structures solved by 
x-ray crystallography, we use a
dataset of $221$ high-resolution protein structures collected by Wang
and Dunbrack~\cite{xtal:WangBioinformatics2003}. Finally, we created a
dataset of structures that have been solved multiple times by x-ray
crystallography, with resolution of $2$\AA~ or less and the same crystal
forms and space groups, to allow us to assess structural variations
that arise from thermal fluctuations.

\begin{figure*}[t]
    \centering
    \includegraphics[height=0.2\textheight]{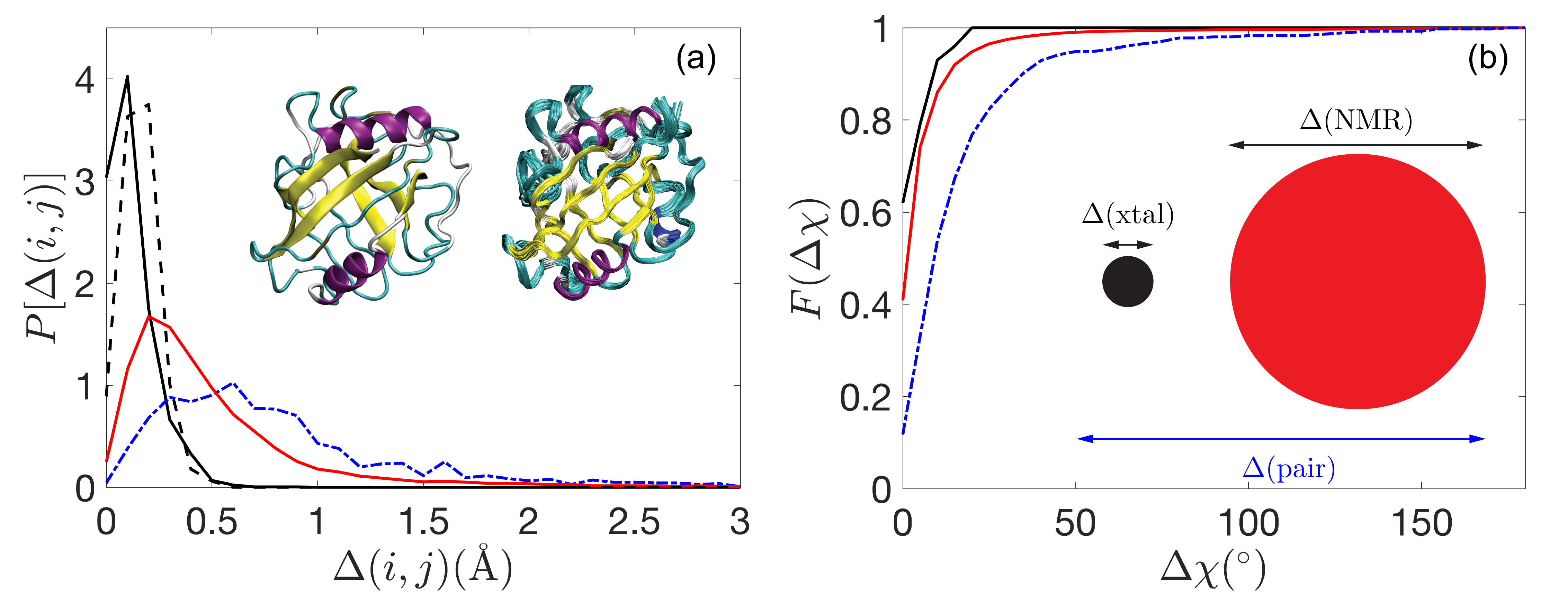}
\caption{(a) Probability distributions $P(\Delta(i,j))$ of the
root-mean-square deviations (RMSD) in the positions of the
C$_{\alpha}$ atoms (in~\AA) for core residues in duplicate x-ray
crystal structures (solid black line), in the NMR model ensemble for each
structure (solid red line), and in paired x-ray crystal and NMR
structures (dot-dashed blue line).  We also plot the distribution
for $\Delta = \sqrt{3B/8\pi^2}$ from the B-factor for core
C$_{\alpha}$ atoms in the duplicate x-ray crystal structures (dashed black
line). The inset shows an example of one of the proteins in the
paired x-ray crystal and NMR structure dataset, with the x-ray
crystal structure on the left and the bundle of $20$ NMR structures on
the right (PDB codes 3K0M and 1OCA, respectively). The
$\alpha$-helices are colored purple, the $\beta$-sheets are yellow,
and the loops are gray. (b) The fraction of core amino acids
$F(\Delta \chi)$ with root-mean-square deviations of the side chain
dihedral angles less than $\Delta \chi$ (in degrees) for the
pairwise comparisons in (a).  The inset is a schematic in two
dimensions of the high-dimensional volume in configuration space
that the C$_{\alpha}$ atoms in core residues in x-ray crystal structures and NMR
ensembles sample. X-ray crystal structures sample a smaller region
than NMR ensembles, but the distance \textit{between} these regions
of configuration space is larger than the fluctuations of both the
x-ray crystal and NMR structures.  The relative lengths of the arrows are drawn to scale,
with $\langle \Delta \rangle \approx 0.1$, $0.5$, and $0.8$ \AA~for
the x-ray duplicates, NMR models, and pairs of x-ray crystal and NMR
structures, respectively.}
\label{fig:fluctuations}
\end{figure*}

We find that the root-mean-square deviations (RMSD) of the positions
of core C$_\alpha$ atoms within an NMR `bundle’ is greater than the
RMSD of core C$_\alpha$ atoms of the set of protein crystal structures that have been solved multiple times. Also, the difference between an x-ray crystal
structure and each structure in the NMR ‘bundle’ is greater than the
spread within the NMR bundle. To gain deeper insight into these
differences, we first analyzed side chain repacking of core residues
in structures determined by both NMR and x-ray crystallography, using
the hard-sphere plus stereochemical constraint model for both.  We have
shown in previous work that the ability to accurately predict the
placement of sidechains of core residues is correlated with the high
packing fraction $\langle \phi \rangle \approx 0.55$ found in protein
cores. We are able to predict the placement of side chains of core
residues to above $90\%$ accuracy to within $30^{\circ}$ in structures
determined by either x-ray crystallography or NMR, which is indicative
of dense packing. When we explicitly calculate the packing fraction of
core residues in protein structures determined by x-ray
crystallography versus NMR, we find that that the cores of NMR
structures are more tightly packed than the cores of x-ray crystal
structures. 

To further explore the physical basis for these observations, we
generated jammed packings of amino-acid-shaped particles
computationally, and explored the extent to which we can tune their
packing fraction using protocols with different degrees of
themalization. We find that depending on the thermalization protocol
we use, we can match the packing fraction to that which we observe in
structures determined by x-ray crystallography and by
NMR. Specifically, the packing fraction of amino acid-shaped particles
we observe in the athermal limit corresponds to the packing fraction
we find in the cores of protein crystal structures, whereas the
packing fraction we observe in the cores of structures determined by
NMR is higher, but less than the packing fraction achieved from a high degree of thermalization. Thus, the core packing fraction we observe for protein
structures determined by x-ray crystallography and NMR are both physically
reasonable, and we speculate that the higher packing fraction we
observe for structures determined by NMR reflects the different
conditions under which NMR structures are determined.

\begin{figure*}
    \centering
    \includegraphics[height=0.175\textheight]{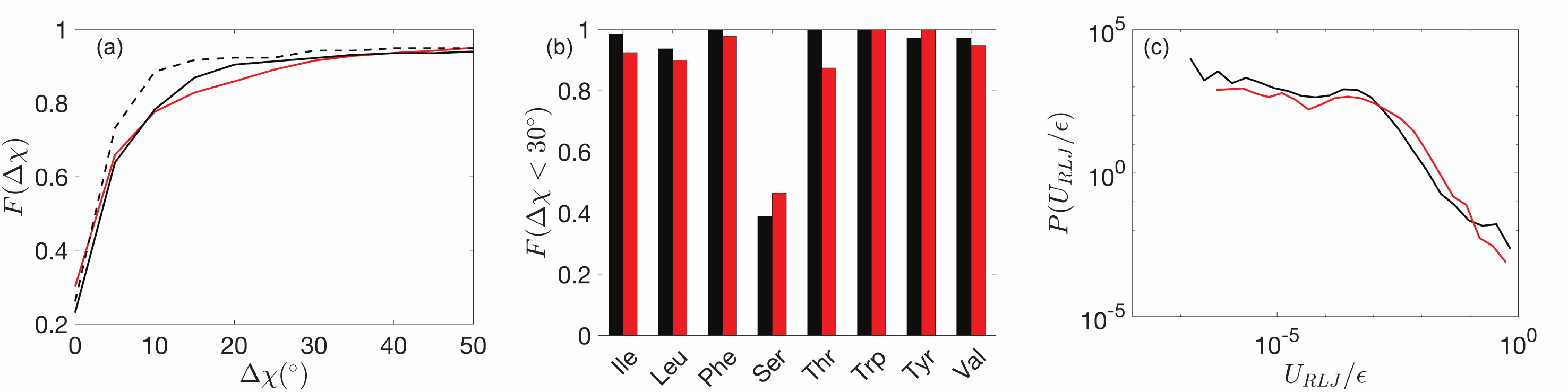}
    \caption{(a) Fraction of side chain conformations of core residues 
with predictions from the hard-sphere plus stereochemical constraint model 
that deviate from the 
experimentally observed values by less than $\Delta \chi$ (in degrees) 
in the dataset of x-ray crystal (black line) and NMR (red line) structure 
pairs, and the Dunbrack 1.0 dataset of $221$ high resolution x-ray 
crystal structures~\citep{xtal:WangBioinformatics2003,xtal:WangNucleicAcidsRes2005}. (b) Fraction of core hydrophobic side chains, grouped by residue type, 
that can be predicted to within $30^\circ$ of the corresponding experimental 
structure using the hard-sphere plus stereochemical constraint model for
xray (black bars) and NMR structures (red bars). (c) Distribution of 
the overlap potential energy $U_{RLJ}/\epsilon$, calculated using Eq.~\eqref{eq:rljEnergyDefintion} for core residues in the x-ray crystal (black line) and NMR structures
(red line) in the paired dataset.}
    \label{fig:repacking}
\end{figure*}

We first compare pairs of structures, determined by x-ray
crystallography and NMR, by quantifying the root-mean-square deviation (RMSD) 
of the C$_\alpha$ positions of a given set of residues
defined by their sequence location on two structures $i$ and $j$:
\begin{equation}\label{eq:rmsdDefinition}
    \Delta(i,j) = \sqrt{\frac{1}{N_S}\sum_{\mu=1}^{N_s}\qty(\vec{c}_{\mu,j} - \vec{c}_{\mu,i})^2},
\end{equation}
where $\vec{c}_{\mu,i}$ is the position of the C$_\alpha$ atom on
residue $\mu$ in structure $i$, and $N_S$ is the number of residues
being compared. For the NMR datasets, $i$ and $j$ represent
each model within a bundle and for the x-ray crystal duplicate
dataset, $i$ and $j$ represent each of the duplicates. We define core
residues as residues with small ($< 10^{-3}$) relative
solvent-accessible surface area (rSASA), as defined in Eq. (1) in the
SI. In Fig.~\ref{fig:fluctuations} (a), we compare the distributions
$P(\Delta(i,j))$ of RMSD values of core residues in x-ray crystal
structure duplicates and RMSD values of core residues in NMR
bundles. We show that the fluctuations among x-ray crystal structure
duplicates are consistent with B-factor fluctuations of the C$_\alpha$
positions of core residues, which are given by $\Delta =
\sqrt{3B/8\pi^2}$. We also compare x-ray crystal and NMR structures
for the same proteins by calculating the RMSD between
C$_{\alpha}$ atoms of core residues.

We also calculate the side chain dihedral angle fluctuations $\Delta \chi$
for the same pairs of structures; we define $\Delta \chi(\mu|i,j)$ as
the distance between the sidechain conformations of residue $\mu$
within structures $i$ and $j$, i.e.
\begin{equation}\label{eq:deltaChiDefinition}
    \Delta\chi(\mu|i,j) = \sqrt{\qty(\vec{\chi}_{\mu,j}-\vec{\chi}_{\mu,i})^2}.
\end{equation}
where $\vec{\chi}_{\mu,i}$ is the set of dihedral angles
$(\chi_1,\ldots,\chi_m)$ for residue $\mu$ on structure $i$. Note that in
Fig. \ref{fig:fluctuations}(b), we measure $\Delta \chi$ between two
experimental structures of the same protein, whereas in
Fig. \ref{fig:repacking} (a) and (b) we measure $\Delta\chi$ between an
experimental structure and a prediction using the hard-sphere plus 
stereochemical constraint model. 

In Fig.~\ref{fig:fluctuations}, we show that the conformations of both
the backbone and sidechains of core residues fluctuate less in x-ray
crystal structures compared to the conformations within an NMR bundle,
but that the fluctuations within an NMR bundle are smaller than those
\textit{between} the x-ray crystal and NMR structure
pairs~\citep{comp:GarbuzynskiyPSFB2005,comp:CarugoOpenBiochemistry2010,comp:EverettProteinScience2016}.
The inset to Fig.~\ref{fig:fluctuations} (b) illustrates the
proportion of configuration space sampled for structures solved by
both NMR and x-ray crystallography. Structures determined by x-ray
crystallography sample states in a relatively small volume of
configuration space compared to that sampled by structures in an NMR
bundle. Moreover, these two ensembles are separated by a
characteristic distance that is larger than the scale of fluctuations
in either ensemble.

To put these structural differences in context, we also analyze
fluctuations in a database of pairs of x-ray crystal structures of
wild-type proteins and the same protein with a single core mutation
and also high-scoring submissions from a recent Critical Assessment of
Protein Structure Prediction (CASP)
competition~\citep{prediction:MoultPSFB2018}. We find that the scale
of the fluctuations of single-site core mutants relative to wildtype
structures is similar to that observed in x-ray crystal structure
duplicates. In contrast, submissions to CASP12 exhibit much larger
fluctuations. Because the CASP12 submissions are computational
predictions, not experimentally determined structures, one would
expect large structural fluctuations among the different CASP12
submissions. The scale of the structural fluctuations among the CASP12
submissions is also larger than those between structures of the same
protein determined by x-ray crystallography or NMR. In the SI, we
report additional measures of structural fluctuations, such as
fluctuations in the identities of residues that make up the core.

\begin{figure}[t]
    \centering
    \includegraphics[width=0.475\textwidth]{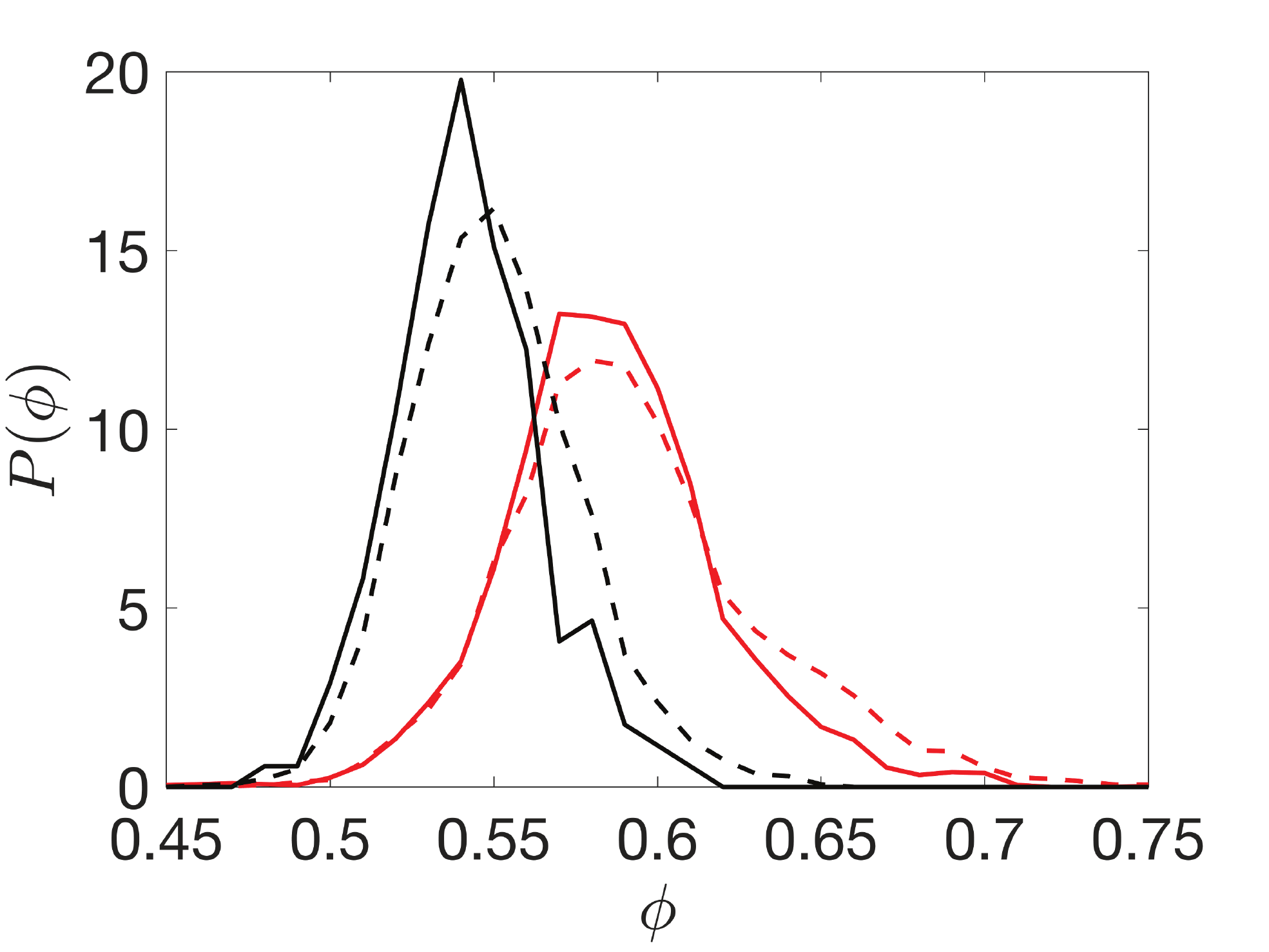}
\caption{Distribution $P(\phi)$ of the packing fraction of core residues 
in the Dunbrack $1.0$ dataset of high-resolution x-ray crystal structures 
(black dashed line), the dataset of high-resolution NMR structures 
for which there is not a corresponding x-ray crystal structure 
(red dashed line), and x-ray crystal structures (black solid line) and NMR 
structures (red solid) from the paired dataset.} 
\label{fig:packingFractionDists}
\end{figure}

To investigate the possible origin of the differences between
structures determined by x-ray crystallography and NMR, we first
investigated if these differences are due to the physical forces
governing side chain placement in the core.  In previous work, we
showed that the hard-sphere plus stereochemical constraint model can
uniquely specify the sidechain dihedral angles of hydrophobic residues
in the cores of protein crystal
structures~\citep{subgroup:ZhouPSFB2014,subgroup:CaballeroPEDS2016,subgroup:GainesPEDS2017,subgroup:GainesProteins2018}. One
potential source of the differences in the fluctuations measured in
NMR and crystal structure cores could be that the cores in NMR
structures are less well-resolved, and the sidechains are poorly
placed due to insufficient information to uniquely define their
conformations. Such methodological inaccuracies have been suggested by
previous studies, where computational refinement moves the NMR
backbone and sidechain dihedral angles towards values similar to those
of x-ray crystal
structures~\citep{comp:GarbuzynskiyPSFB2005,comp:SchneiderPSFB2009,comp:MaoJACS2014,comp:LemanPSFB2018}. However,
as shown in Fig.~\ref{fig:repacking}, we find that with our
high-quality dataset of NMR structures we can repack hydrophobic side
chains of core residues just as accurately as we can repack the same
side chains in high-resolution x-ray crystal structures. The repacking
protocol is outlined in detail in the SI, but briefly, we define the
probability that a given residue $\mu$ has adopted a given side chain
confirmation $\vec{\chi}_{\mu}$ as $P(\vec{\chi}_{\mu}) \propto \exp[-\beta
  U_{\text{RLJ}}(\vec{\chi}_{\mu})]$, where
\begin{equation}\label{eq:rljEnergyDefintion}
    U_{\text{RLJ}}(\vec{\chi}_{\mu}) = \sum_{\nu=1}^N\sum_{i,j} \frac{\epsilon}{72}\qty[1-\qty(\frac{\sigma_{ij}^{\mu\nu}}{r^{\mu\nu}_{ij}})^6]^2\Theta\qty(\sigma_{ij}^{\mu\nu}-r_{ij}^{\mu\nu})
\end{equation}
is the purely repulsive Lennard-Jones potential energy of residue
$\mu$, evaluated as a sum over all non-bonded atomic interactions.
$r_{ij}^{\mu\nu}$ is the distance between atoms $i$ and $j$ on
residues $\mu$ and $\nu$,
$\sigma_{ij}^{\mu\nu} = \qty(\sigma_i^\mu + \sigma_j^\nu)/2$, and
$\sigma_i^\mu$ is the diameter of atom $i$ on residue $\mu$. 

However, when we investigate the packing fraction $\phi$ of core
residues for x-ray crystal and NMR structures, we find important
differences.  In Fig.~\ref{fig:packingFractionDists}, we plot the
probability distribution $P(\phi)$ of the packing fraction for core
residues in x-ray crystal and NMR structures. The average packing
fraction of core residues in the protein structures in the datasets
determined by x-ray crystallography is $\langle \phi \rangle = 0.55
\pm 0.01$, a value that is consistent with our previous results for
the packing fraction of core residues in globular and transmembrane
protein cores and the cores of protein-protein interfaces solved by
x-ray crystallography~\citep{subgroup:GainesProteins2018}.  For core
residues of protein structures in the NMR database, the average
packing fraction is higher with $\langle \phi \rangle =0.59\pm
0.02$. We believe that this is the first time that such a difference
in the packing fraction of core residues in high-quality protein
structures determined by both x-ray crystallography and NMR has been
reported.

We were concerned that the higher packing fraction of core residues in 
protein structures determined by NMR could be an artifact, for example,
the result of improperly-placed sidechains that overlap with
neighboring residues, which would artificially increase the observed
packing fraction.  With this concern in mind, we calculated and
compared the potential energy of atomic overlaps of non-bonded atoms $U_{RLJ}$
for the structures determined by x-ray crystallography and
NMR. However, we found that structures determined by either method
have virtually identical overlap energies as shown in Fig.~\ref{fig:repacking}
(c), and we therefore ruled out this potential cause. The difference
in the packing fraction of core residues was at first surprising,
because our previous studies showed that the cores of x-ray crystal
structures pack as densely as jammed packings of purely-repulsive
amino-acid-shaped particles without backbone
constraints generated using a protocol of successive compressions 
followed by energy minimization~\citep{subgroup:GainesPRE2016,subgroup:TreadoPRE2019}. 

We therefore revisited the protocol with which we prepared jammed
packings of amino-acid-shaped
particles~\citep{subgroup:GainesPRE2016,subgroup:TreadoPRE2019}. In
our previous work, packings were prepared using an ``athermal''
protocol, where kinetic energy was drained rapidly from the system
during the packing preparation. For the athermal protocol, amino acids
were initialized in a cubic simulation box at a small initial packing
fraction $\phi _0$ and compressed by small increments $\Delta \phi$
with each followed by energy minimization. (See SI for additional
details.) Because the amino-acid-shaped particles were not allowed to
translate and rotate significantly between each compression step, the
jammed packings at $\langle \phi \rangle \approx 0.55$ were obtained
at the first metastable jammed state that the protocol
encountered. Thus,
the packing fractions that can be achieved by amino-acid-shaped
particles is protocol-dependent. We therefore investigated more thermalized
protocols to generate jammed packings of amino-acid-shaped particles.

\begin{figure}[t]
    \centering
    \includegraphics[width=0.45\textwidth]{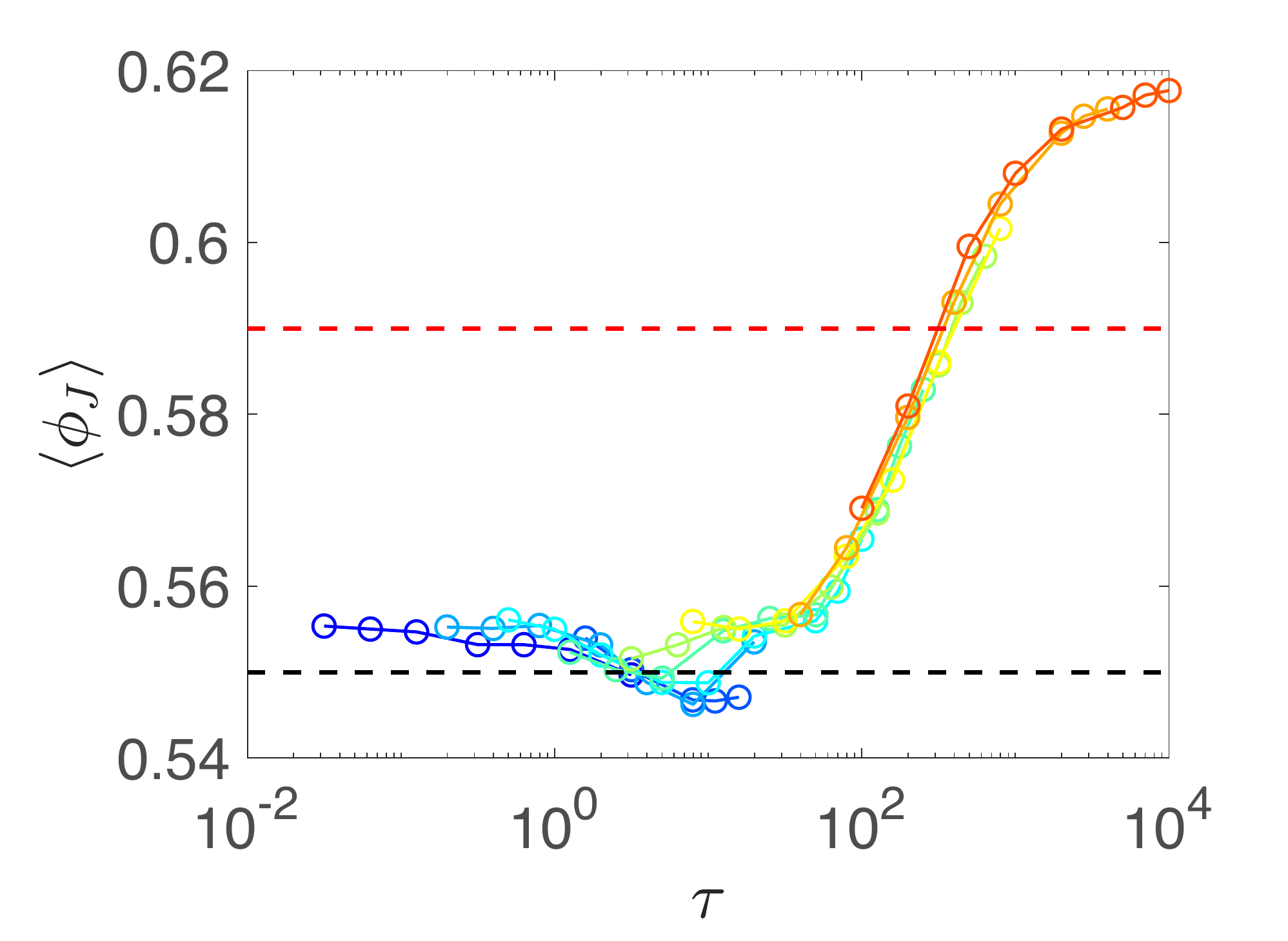}
    \caption{
        The ensemble-averaged packing fraction $\langle \phi_J \rangle$ of 
jammed packings of amino-acid-shaped particles versus the dimensionless 
thermalization timescale $\tau$ for a system with $N = 16$ particles. 
Different colors represent simulations with different dimensionless 
temperatures $k_B T/\epsilon$, logarithmically spaced from $10^{-7}$ (blue) 
to $1$ (red). The dashed black line at $\langle \phi_J \rangle = 0.55$ 
is the average packing fraction of core residues in x-ray crystal structures, 
and the dashed red line at $\langle \phi_J \rangle= 0.59$ is the average 
packing fraction of core residues in NMR structures.}
    \label{fig:simulation}
\end{figure}

We chose a family of quasi-annealing packing-generation protocols. We
initialize the system in a dilute configuration, and compress
the system in small increments $\Delta \phi$ between periods of
molecular dynamics simulations in the canonical ensemble for a time
period $t_{MD}$ at thermal energy $k_BT$. (See SI for details.)  We
find that temperature only acts to renormalize the simulation time
window $t_{MD}$, i.e. a longer simulation run at a lower temperature
will yield the same results as a shorter simulation run at a higher
temperature. Thus, there is another time scale associated with the
quasi-annealing protocol, $t_{\text{QA}} = c(T)t^*$, where $c(T)$ is a
dimensionless quantity that depends on temperature, $t^* =
\sqrt{m_R\sigma_R^2/\epsilon}$, and $m_R$ and $\sigma_R$ are the mass
and diameter of the smallest residue. We find that plotting the
packing fraction versus $\tau$, where
\begin{equation}
    \tau = t_{\text{MD}}/t_{\text{QA}} =n\qty(\frac{k_B T}{\epsilon})^\alpha,
\end{equation}
collapses the data for different temperatures and time periods onto a
single curve as shown in Fig.~\ref{fig:simulation}. The exponent 
$\alpha = 0.4 \pm 0.01$ and $n$ is the number of time steps
between compression increments. 

Two limits of packing fractions emerge over the broad range of
quasi-annealing protocols we tested; an athermal limit, which
corresponds to packing fractions one finds in the cores of x-ray
crystal structures
~\citep{subgroup:GainesProteins2018,subgroup:GainesPRE2016}, and the
completely thermalized limit, which can reach average packing
fractions $\langle \phi \rangle \approx 0.62$. The packing fractions
observed in the cores of protein structures solved by NMR fall between
these two extremes with $\langle \phi \rangle = 0.59$.  The states at
exceedingly high packing fractions exist only in the limit of
extremely long annealing times. The results of simulations using different protocols are consistent with the differences observed in the cores of protein structures solved by x-ray crystallography and NMR. The fact that thermalized packing protocols can
yield NMR-like packing fractions, and that athermal protocols generate
x-ray crystal-like packing fractions, suggests that native-state fluctuations
are distinct for these two methods.

A previous study that also compared protein structures determined by
x-ray crystallography and NMR suggested that the crystal environment
restricts dynamical fluctuations, whereas bundles of NMR structures in
solution contain the full dynamics one would expect from elastic
network models for proteins~\citep{comp:YangStructure2007}. The work
we present here provides significant further evidence to support this
conclusion, but whether the differences are due to crystalline
contacts~\citep{comp:YangStructure2007,comp:CarugoOpenBiochemistry2010,xtal:HallePNAS2004}
or the different temperatures at which the protein structures are
characterized~\citep{xtal:FraserPNAS2011,sim:HuNaturePhysics2015}
remains to be determined.

\section*{Author Contributions}

Z.M. and J.D.T. contributed equally to this article. Z.M., J.D.T.,
Z.A.L., L.R., and C.S.O. designed the research. Z.M. compiled the
dataset of protein structures, J.D.T. carried out simulations, and
Z.M., J.D.T., Z.A.L., L.R. and C.S.O. interpreted the results. Z.M.,
J.D.T., C.S.O. and L.R. wrote the article.

\section*{Acknowledgments}
The authors acknowledge support from NIH training Grant
No. T32EB019941 (J.D.T.), the Raymond and Beverly Sackler
Institute for Biological, Physical, and Engineering Sciences
(Z.M.), and NSF Grant No. PHY-1522467 (C.S.O.).
This work also benefited from the facilities and staff of the
Yale University Faculty of Arts and Sciences High Performance
Computing Center. We thank Pat Loria and Peter Moore for helpful 
discussions. 

\bibliography{nmr}

\end{document}